\documentclass[numberedappendix]{emulateapj}
\newcommand{\myemail}{toshiki.saito@nao.ac.jp}
\usepackage[breaklinks,colorlinks,urlcolor=blue,citecolor=blue,linkcolor=blue]{hyperref}

\shorttitle{Multiple CH$_3$OH Line Imaging toward VV~114}
\shortauthors{Toshiki Saito et al.}

\begin{document}

\title{Merger-induced Shocks in the Nearby LIRG VV~114 through Methanol Observations with ALMA}

\author{
   Toshiki \textsc{Saito}\altaffilmark{1,2},
   %et al.\if0
   Daisuke \textsc{Iono}\altaffilmark{2,3},
   Daniel \textsc{Espada}\altaffilmark{2,3},
   Kouichiro \textsc{Nakanishi}\altaffilmark{2,3},
   Junko \textsc{Ueda}\altaffilmark{4},
   Hajime \textsc{Sugai}\altaffilmark{5},
   Shuro \textsc{Takano}\altaffilmark{6},
   Min \textsc{S. Yun}\altaffilmark{7},
   Masatoshi \textsc{Imanishi}\altaffilmark{8},
   Satoshi \textsc{Ohashi}\altaffilmark{1,2},
   Minju \textsc{Lee}\altaffilmark{1,2},
   Yoshiaki \textsc{Hagiwara}\altaffilmark{9},
   Kentaro \textsc{Motohara}\altaffilmark{10},
   and
   Ryohei \textsc{Kawabe}\altaffilmark{1,2,3}%\fi
}

\email{\myemail}
\altaffiltext{1}{Department of Astronomy, The University of Tokyo, 7-3-1 Hongo, Bunkyo-ku, Tokyo 113-0033, Japan}\if0
\altaffiltext{2}{National Astronomical Observatory of Japan, 2-21-1 Osawa, Mitaka, Tokyo, 181-0015, Japan}
\altaffiltext{3}{The Graduate University for Advanced Studies (SOKENDAI), 2-21-1 Osawa, Mitaka, Tokyo 181-0015, Japan}
\altaffiltext{4}{Harvard-Smithsonian Center for Astrophysics, 60 Garden Street, Cambridge, MA 02138, USA}
\altaffiltext{5}{Kavli Institute for the Physics and Mathematics of the Universe (WPI), The University of Tokyo, 5-1-5 Kashiwanoha, Kashiwa, 277-8583, Chiba}
\altaffiltext{6}{Physics Department, College of Engineering, Nihon University, 1 Nakagawara, Tokusada, Tamura, Koriyama, Fukushima, 963-8642, Japan}
\altaffiltext{7}{Department of Astronomy, University of Massachusetts, Amherst, MA 01003, USA}
\altaffiltext{8}{Subaru Telescope, 650 North A’ohoku Place, Hilo, Hawaii, 96720, USA}
\altaffiltext{9}{Natural Science Laboratory, Toyo University, 5-28-20, Hakusan, Bunkyo-ku, Tokyo 112-8606, Japan}
\altaffiltext{10}{Institute of Astronomy, The University of Tokyo, 2-21-1 Osawa, Mitaka, Tokyo 181-0015, Japan}\fi

\received{August 30, 2016}
\revised{October 10, 2016}
\accepted{November, 4, 2016}

\begin{abstract}
We report the detection of two CH$_3$OH lines ($J_K$ = 2$_K$--1$_K$ and 3$_K$--2$_K$) between the progenitor's disks (``Overlap'') of the mid-stage merging galaxy VV~114 obtained using the Atacama Large Millimeter/submillimeter Array (ALMA) Band~3 and Band~4.
The detected CH$_3$OH emission show an extended filamentary structure ($\sim$ 3~kpc) across the progenitor's disks with relatively large velocity width (FWZI $\sim$ 150 km s$^{-1}$).
The emission is only significant in the ``overlap" and not detected in the two merging nuclei.
Assuming optically-thin emission and local thermodynamic equilibrium (LTE), we found the CH$_3$OH column density relative to H$_2$ ($X_{\rm CH_3OH}$) peaks at the ``Overlap" ($\sim$ 8 $\times$ 10$^{-9}$), which is almost an order of magnitude larger than that at the eastern nucleus.
We suggest that kpc-scale shocks driven by galaxy-galaxy collision may play an important role to enhance the CH$_3$OH abundance at the ``Overlap".
This scenario is consistent with that shock-induced large velocity dispersion components of ionized gas have been detected in optical wavelength at the same region.
Conversely, low $X_{\rm CH_3OH}$ at the nuclear regions might be attributed to the strong photodissociation by nuclear starbursts and/or putative active galactic nucleus (AGN), or inefficient production of CH$_3$OH on dust grains due to initial high temperature conditions (i.e., desorption of the precursor molecule, CO, into gas-phase before forming CH$_3$OH on dust grains).
These ALMA observations demonstrate that CH$_3$OH is a unique tool to address kpc-scale shock-induced gas dynamics and star formation in merging galaxies.

\end{abstract}


\section{INTRODUCTION}
Bright thermal rotational transitions of methanol (CH$_3$OH) are often used as a tracer of extragalactic shocks, which is established by a large number of unbiased wide-band mm/sub-mm molecular line surveys toward galaxies over the last decade \citep[][and references therein]{Takano14}.
Extragalactic CH$_3$OH observations found that some of the galaxies have $X_{\rm CH_3OH}$ larger than $\sim$ 10$^{-8}$ at the nuclei, arms, and bars with $\sim$ 100~pc resolution \citep[e.g., IC~342;][]{Meier&Turner05}.
Purely gas-phase chemistry cannot explain these observational evidences, because
the formation process of CH$_3$OH in gas-phase is not efficient to produce $X_{\rm CH_3OH}$ greater than $\sim$ (1-3) $\times$ 10$^{-9}$ \citep{Lee96}.
Alternatively, high CH$_3$OH abundance is believed to arise from a series of hydrogenations of CO on dust grain surfaces under a low-temperature ($\sim$ 10 K) condition \citep{Watanabe03}, because interstellar icy mantles are rich in CH$_3$OH \citep[$\sim$ 10$^{-6}$; e.g.,][]{Schutte91}.
After the production on dust, it requires high temperature \citep[i.e., hot-core chemistry; e.g.,][]{Garrod08} or energetic heating \citep[i.e., shock chemistry; e.g.,][]{Viti11} mechanisms to heat the dust and then sublime CH$_3$OH into gas-phase.
On the other hand, CH$_3$OH molecules are easily destroyed by UV radiation due to starburst or AGN without shielding \citep[$A_V$ $\sim$ 5;][]{Martin09}.
Thus, CH$_3$OH requires hot-core like or shocked ISM without strong UV radiation field to achieve an observable abundance.
When hot-core like environment is excluded, CH$_3$OH lines become an excellent extragalactic shock tracer in molecular ISM.

However, most of the previous studies had mainly focused on nearby, bright galaxies or their nuclear regions so far \citep{Henkel87,Meier&Turner05,Martin06a,Martin06b,Usero06,Garcia-Burillo10,Aladro11,Costagliola11,Martin11,Meier&Turner12,Davis13,Meier&Turner14,Takano14,Watanabe14,Aladro15,Costagliola15,Nakajima15,Galametz16,Nishimura16}.
In this paper, we present multiple CH$_3$OH line observations toward the nearby merging galaxy VV~114, which are follow-up observations of our previous CH$_3$OH~(2$_K$--1$_K$) detection at Overlap \citep{Saito15}, in order to test kpc-scale shocks in molecular ISM due to a gas-rich galaxy-galaxy collision.
%Those data are obtained as a part of spectral scan mode observations of Band~3 and Band~4 of ALMA.

VV~114 is one of the local luminous infrared galaxies \citep[$D_{\rm L}$ = 87 Mpc, $L_{\rm IR}$ = 10$^{11.69}$ $L_{\odot}$;][]{Armus09}.
It is considered to be a mid-stage gas-rich major merger with the nuclear separation of $\sim$ 6~kpc.
The system has a (molecular and ionized) gaseous and dust filamentary structure ($\sim$ 4 kpc) across the galaxy disks \citep{Yun94,Frayer99,Iono04,Wilson08,Rich11,Iono13,Sliwa13,Saito15,Tateuchi15}, which is not found in optical, FUV, and X-ray images \citep{Scoville00,Goldader02,Grimes06}.

Our previous ALMA observations \citep{Iono13,Saito15} revealed that the filament has multiple star-forming dense gas clumps at Overlap, one of which only has a CH$_3$OH~(2$_K$--1$_K$) peak.
In contrast, two massive clumps in the eastern nucleus, which harbor compact starburst (``SB") and putative hard X-ray AGN (``AGN"), are bright in 
CN, HCN, and HCO$^+$ lines.
This indicates that VV~114 has kpc-scale chemical variations along the filament.
However, since we were only able to observe ten molecular lines, the chemical composition in the filament is not fully understood, and thus we followed up VV~114 with ALMA spectral scan mode.

We assumed $H_0$ = 70~km~s$^{-1}$~Mpc$^{-1}$, $\Omega_{\rm M}$ = 0.3, $\Omega_{\rm \Lambda}$ = 0.7 (1\arcsec = 420~pc) throughout this paper.

\section{OBSERVATIONS AND DATA OVERVIEW}
We observed VV~114 using Band~3 and Band~4 of ALMA with the spectral scan mode \citep{Mathys13} as a cycle~2 program (ID: 2013.1.01057.S).
The correlator was configured to cover 84 - 111~GHz and 127 - 154~GHz using eight tunings.
In this paper, we present the tunings containing the multiple CH$_3$OH~(2$_K$--1$_K$) ($\nu_{{\rm obs}}$ $\sim$ 94.84~GHz) and CH$_3$OH~(3$_K$--2$_K$) ($\nu_{{\rm obs}}$ $\sim$ 142.24~GHz) lines.
We observed blended sets of thermal rotational transitions of CH$_3$OH.
Detected transitions are listed in Table~\ref{table_1}.
Other transitions (e.g., $J_K$ = 0$_0$--1$_{-1}$) are not robuestly detected.
Since such non-detections do not put meaningful upper limits in the analysis shown later, we ignore them in this paper.
We will discuss some non-detected molecular lines in Saito et al. in preparation.
The CH$_3$OH~(2$_K$--1$_K$) data (Band~3) were obtained on 2014 July 3, 2014 July 4, and 2015 June 11 with the double-sideband system temperature ($T_{\rm sys}$) of 39 - 103~K, whereas the CH$_3$OH~(3$_K$--2$_K$) data (Band~4) was obtained on 2015 May 25 with $T_{\rm sys}$ of 47 - 124~K.
Thirty-one to thirty-eight (thirty-six) 12~m antennas with the projected baseline length of 19 - 778~m (21 - 539~m) were assigned for the Band~3 (Band~4) observations.
Each tuning has four spectral windows to cover each sideband.
The spectral window has a bandwidth of 1.875~GHz with 1.938~MHz resolution, while they were binned together to create better signal-to-noise (S/N) data cubes.
The total on-source time of the Band~3 and Band~4 observations are 47.4~min. and 20.0~min., respectively.
Neptune and Uranus were used for the flux calibration of the both bands.
J0137-2430 and J2258-2758 were used for the bandpass calibration of the both bands.
Either of one of J0116-2052 and J0110-0741 was used for the phase calibration of Band~3, whereas J0110-0741 was used for that of Band~4.

The data reduction, calibration, and imaging were carried out using {\tt CASA} \citep{McMullin07}.
All maps are reconstructed with the natural weighting to minimize noise rms levels.
We made the data cubes with a velocity resolution of 50 km s$^{-1}$ ($\sim$ 14.4~MHz) for CH$_3$OH~(2$_K$--1$_K$) and 20 km s$^{-1}$ ($\sim$ 9.8~MHz) for CH$_3$OH~(3$_K$--2$_K$).
Before imaging, we combined the cycle~0 CH$_3$OH~(2$_K$--1$_K$) data
(ID: 2011.0.00467.S)
in order to increase the sensitivity, and continuum emission was subtracted in the $uv$-plane.
We note that we re-calibrated the cycle~0 data using Butler-JPL-Horizons 2012 model.
Other imaging properties are listed in Table~\ref{table_1}, and will be introduced in detail by Saito et al. in preparation.
The systematic error of absolute flux scaling factor using a solar system object is 5\% for Band~3 and Band~4 \citep{Lundgren13}.
The flux densities of the bandpass and phase calibrators were in good agreement with measurements provided by the ALMA Calibrator Source Catalogue\footnote{https://almascience.nrao.edu/sc/}.
We ignore the difference of missing flux effect between Band~3 and Band~4, since the maximum recoverable scale of the assigned configuration for the Band~3 and Band~4 observations (18\arcsec and 13\arcsec, respectively) is comparable or larger than the filament of VV~114 detected in $^{13}$CO~(1--0) and dust continuum \citep{Saito15}.

\section{RESULTS}
\begin{figure*}
\begin{center}
\includegraphics[width=18cm]{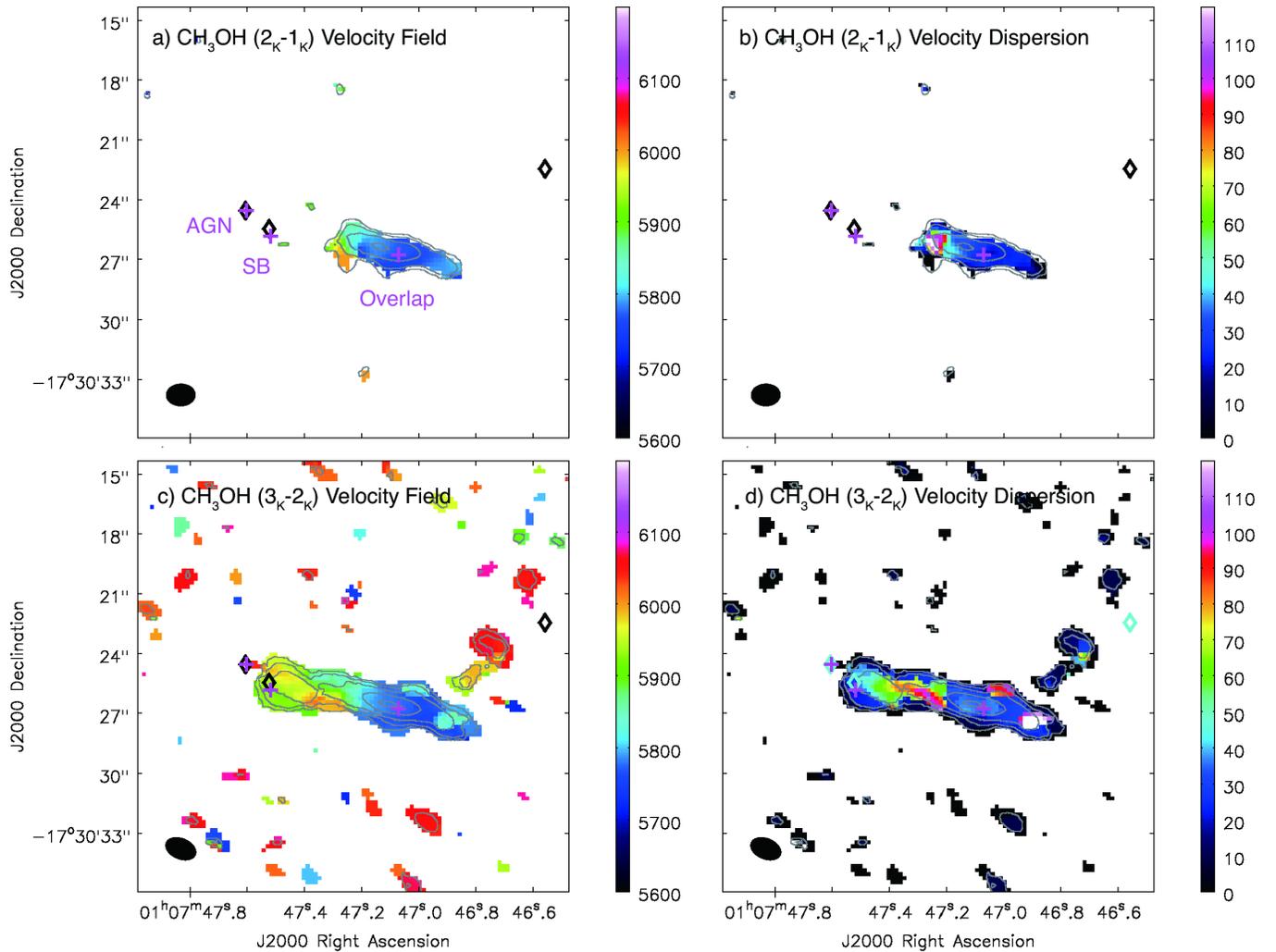}
\caption{(a) Integrated intensity contour of CH$_3$OH~(2$_K$--1$_K$) overlaid on the velocity field. The contours are 0.32 $\times$ (0.16, 0.32, 0.64, and 0.96) Jy beam$^{-1}$ km s$^{-1}$. (b) Same as (a) but for CH$_3$OH~(3$_K$--2$_K$). The contours are 0.67 $\times$ (0.08, 0.16, 0.32, 0.64, and 0.96) Jy beam$^{-1}$ km s$^{-1}$. The crosses show the peak positions of the HCO$^+$~(4--3) emission. AGN, SB, and Overlap correspond to E0, E1, and W0 defined by \citet{Iono13}, respectively. The synthesized beams are shown in the bottom-left corner.  The diamonds show the K$_s$-band stellar nuclei.
}
\label{fig_1a}
\end{center}
\end{figure*}

\begin{figure*}
\begin{center}
\includegraphics[width=18cm]{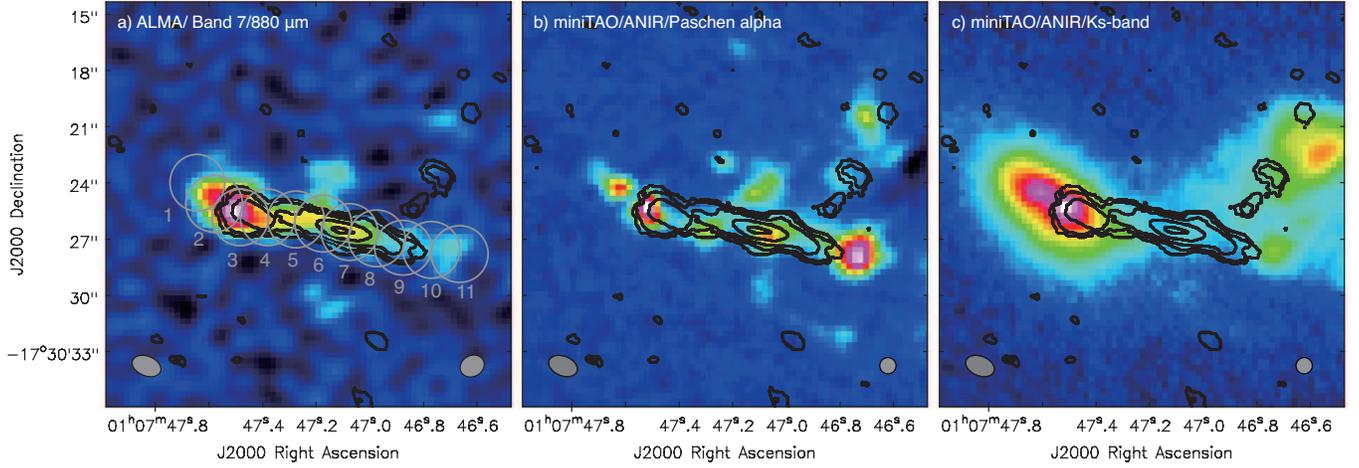}
\caption{Integrated intensity contour of CH$_3$OH~(3$_K$--2$_K$) overlaid on (a) 880~$\mu$m dust emission, (b) Paschen~$\alpha$ emission, and (c) K$s$-band \citep{Saito15,Tateuchi15}. The synthesized beam of the CH$_3$OH~(3$_K$--2$_K$) image is shown in the bottom-left corner.  The beam size of each color image is shown in the bottom-right corner.  The eleven gray circles shown in Figure~\ref{fig_1b}a are used for the photometry along the filament.
}
\label{fig_1b}
\end{center}
\end{figure*}

Images of the integrated intensity, velocity field, and velocity dispersion are shown in Figure~\ref{fig_1a}.
The total CH$_3$OH~(2$_K$--1$_K$) and CH$_3$OH~(3$_K$--2$_K$) integrated intensities are 0.91 $\pm$ 0.11 and 2.31 $\pm$ 0.18 Jy km s$^{-1}$, respectively.
We checked possible blending with other species using the molecular line database Splatalogue\footnote{http://www.splatalogue.net/} and also line detections reported by \citet{Watanabe14}, \citet{Aladro15} and \citet{Costagliola15}, because some extragalactic line surveys reached line confusion limits.
We found no potential bright lines around CH$_3$OH~(2$_K$--1$_K$).
Although there is $c$-C$_3$H$_2$~(3$_{12}$-2$_{21}$) ($\nu_{\rm rest}$ = 145.08961~GHz) around CH$_3$OH~(3$_K$--2$_K$) ($\nu_{\rm rest}$ = 145.09--145.14~GHz), its contribution to the CH$_3$OH~(3$_K$--2$_K$) intensities may be negligible because any other $c$-C$_3$H$_2$ transitions are not robustly detected toward the Overlap region in our Band~3/4 data.
This is consistent with that the starburst galaxy NGC~253, which is thought to be dominated by shocks, shows weaker $c$-C$_3$H$_2$ line emissions than the CH$_3$OH~(2$_K$--1$_K$) line \citep{Aladro15}.
Toward SB (Figure~\ref{fig_1b}), we detected some $c$-C$_3$H$_2$ transitions (e.g., $J_{KaKc}$ = 4$_{04}$--3$_{13}$), which is not stronger than CH$_3$OH~(3$_K$--2$_K$), so CH$_3$OH~(3$_K$--2$_K$) intensity might be slightly overestimated.
However, this does not change our discussion and conclusion (see Section~\ref{LTE}).

Figure~\ref{fig_1a} shows an extended filamentary structure which coincides with previous molecular gas and dust images \citep{Iono13,Saito15}.
The strongest peak is located at Overlap, whereas the emission is only marginally detected in the stellar nuclei
seen in K$s$-band
(Figure~\ref{fig_1b}c).
The global distribution is consistent with $^{13}$CO and HCO$^+$ lines \citep[molecular gas;][]{Iono13,Saito15} and 880~$\mu$m continuum (dust; Figure~\ref{fig_1b}a), although all of them except for CH$_3$OH show strong peaks at AGN and SB.
In contrast, CH$_3$OH does not coincide with Paschen $\alpha$ emission (H$_{\rm II}$ regions and/or ionized gas shocks; Figure~\ref{fig_1b}b).
Ks-band continuum, mainly tracing old stellar component (Figure~\ref{fig_1b}c), has no peaks at Overlap.
This indicates that the CH$_3$OH filament of VV~114 is a relatively young structure compared with the age of the progenitor galaxies, which is likely due to a galaxy-galaxy collision as predicted by numerical simulations of a gas-rich major merger \citep{Saitoh09,Teyssier10}.
These morphological comparisons suggest that CH$_3$OH lines trace a kpc-scale gas structure at the collision interface of the galaxy-galaxy interaction.
To determine physical properties of Overlap, we employ the rotation diagram method \citep{Goldsmith&Langer99} in the next Section.

\section{LTE CALCULATION OF ROTATION TEMPERATURE AND COLUMN DENSITY} \label{LTE}
\begin{figure}
\begin{center}
\includegraphics[width=8cm]{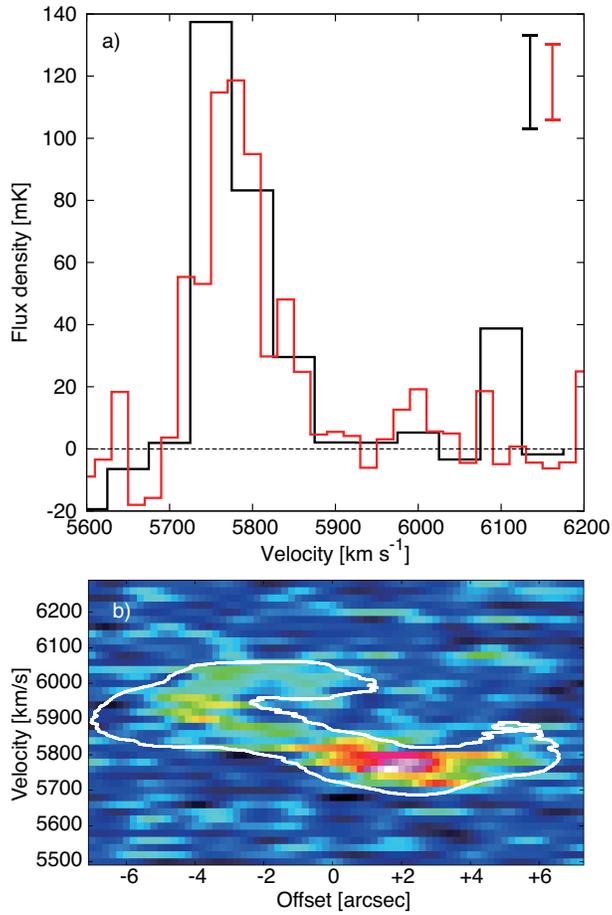}
\caption{(a) Observed spectrum of CH$_3$OH~(2$_K$--1$_K$) (black) and CH$_3$OH~(3$_K$--2$_K$) (red) toward Region 6 (see Figure~\ref{fig_1b}a).  The bars shown in the top right represents a typical statistical error bars associated with the data.  (b) Position-velocity diagram of CH$_3$OH along the filament of VV~114 (position angle = 79.4\degr, length = 14.2\arcsec, and width = 1\farcs8).  The white outline shows the 20$\sigma$ contour of CO~(1--0) \citep{Saito15}.
}
\label{fig_0}
\end{center}
\end{figure}

Assuming LTE and optically thin conditions for the CH$_3$OH lines, we can estimate the column density ($N_{\rm CH_3OH}$) and rotation temperature ($T_{\rm{rot}}$) using the rotation diagram.
However, we need a special treatment in the case of VV~114, because the detected CH$_3$OH emission were blended by multiple $A$-type and $E$-type rotational transitions \citep[e.g.,][]{Rabli10} due to the coarse velocity resolution and observed large velocity widths (FWZI $\sim$ 150~km~s$^{-1}$) as seen in a spectrum toward Region~6 (Figure~\ref{fig_0}a) and a position-velocity diagram along the CH$_3$OH filament (Figure~\ref{fig_0}b).
By assuming that both types of CH$_3$OH have the same beam-averaged column density ($N_{\rm CH_3OH}$), the CH$_3$OH flux can be expressed by using the least squares method with the following equation \citep[e.g.,][]{Martin06a,Watanabe14},
\begin{eqnarray}
W_{\rm \nu} &=& \sum_i\frac{8\pi^3S_i\mu_0^2\nu_iN_{\rm CH_3OH}}{3kQ_{\rm rot}}\left\{1-\frac{\exp(h\nu_i/kT_{\rm rot})-1}{\exp(h\nu_i/kT_{\rm bg})-1}\right\}\nonumber \\
&& \times \exp\left(-\frac{E_{\rm u, i}}{kT_{\rm rot}}\right),
\end{eqnarray}
where $W_{\rm \nu}$ is the flux density, $S_i$ is the line strength, $\mu_0$ is the dipole moment, $\nu_i$ is the frequency of the transition, $Q_{\rm rot}$ is the rotational partition function, $k$ is the Boltzmann constant, $h$ is the Planck constant, $T_{\rm bg}$ is the cosmic microwave background temperature (= 2.73~K), and $E_{\rm u}$ is the upper state energy.
We take the transition parameters necessary for calculating the equation (Table~\ref{table_1}) from Splatalogue\footnotemark[3] and the Cologne Database for Molecular Spectroscopy \citep[CDMS\footnote{http://www.astro.uni-koeln.de/cdms/catalog\#partition};][]{Muller01,Muller05}.

We performed the calculation for eleven apertures of 3\arcsec diameter along the filament shown in Figure~\ref{fig_1b}, and an example for Region~6 is shown in Figure~\ref{fig_2}a.
Parameters used for the calculation and the results are listed in Table~\ref{table_2}.
Each point in Figure~\ref{fig_2}a corresponds to the upper state column density of each unresolved transition, which decomposed by using the best-fit $T_{\rm rot}$.
The average $T_{\rm rot}$ is 7.4 $\pm$ 0.5~K for regions detected in both blended CH$_3$OH sets (i.e., Region~5--8).
For regions without CH$_3$OH~(2$_K$--1$_K$) detection (Region~1--4 and 9--11), we used $T_{\rm rot}$ = 7.4~K to derive $N_{\rm CH_3OH}$.
The average $N_{\rm CH_3OH}$ is (2.3 $\pm$ 1.0) $\times$ 10$^{14}$~cm$^{-2}$, although this is an average of the beam-averaged column densities due to uncorrected beam filling factor ($\eta_{\rm bf}$).
To ignore unknown $\eta_{\rm bf}$ effect, we divided the $N_{\rm CH_3OH}$ by the H$_2$ column density ($N_{\rm CH_3OH}/N_{\rm H_2}$ = $X_{\rm CH_3OH}$) assuming that the $N_{\rm H_2}$ tracer has a same $\eta_{\rm bf}$.
In this paper, we employed the 880~$\mu$m dust continuum map (Figure~\ref{fig_1b}a) to derive molecular gas mass (i.e., $N_{\rm H_2}$) using the formulation described in \citet{Scoville15}.
Assuming a constant dust temperature ($T_{\rm dust}$) of 25~K, we derived $X_{\rm CH_3OH}$ along the filament.
As shown in Figure~\ref{fig_2}b, the $X_{\rm CH_3OH}$ distribution is clearly peaked around Overlap (Region 6--8), whereas the AGN (Region 1) and SB (Region 3) positions show almost an order of magnitude lower values.
The errors shown in Figure~\ref{fig_2}b do not include the systematic uncertainty due to the 880~$\mu$m to $N_{\rm H_2}$ conversion.

The adopted $T_{\rm rot}$ of 7.4~K for Region 1--2 and 9--11 is consistent with the non-detections of the observed CH$_3$OH~(2$_K$--1$_K$) flux, although Region 3 and 4 should be detected when assuming 7.4~K.
This might be attributed to (a) higher $T_{\rm rot}$ than 7.4~K and/or (b) CH$_3$OH~(3$_K$--2$_K$) overestimation due to a contamination from $c$-C$_3$H$_2$~(3$_{12}$-2$_{21}$) emission.
In case (a), since we need to increase $T_{\rm rot}$ to $\sim$ 10~K at least, the derived $N_{\rm CH_3OH}$ will decrease ($\sim$ 20\%), and thus $X_{\rm CH_3OH}$ will decrease.
In case (b), the intrisic CH$_3$OH~(3$_K$--2$_K$) fluxes will decrease 15\% and 30\% at Region 3 and 4, respectively, so that $X_{\rm CH_3OH}$ will decrease.
For both cases, $X_{\rm CH_3OH}$ will decrease by a few tens of \% around the eastern nucleus, so that our discussion and conclusion do not change.
We note that the global $X_{\rm CH_3OH}$ trend is robust even if AGN and SB have higher $T_{\rm rot}$ and $T_{\rm dust}$ conditions, since $X_{\rm CH_3OH}$ only increases $\sim$ 2 times at most when adopting $T_{\rm rot}$ = 7.4--14.8~K and $T_{\rm dust}$ = 25--50~K.
%The result does not change when we use $^{13}$CO to derive $N_{\rm H_2}$.

Comparing with CH$_3$OH observations toward giant molecular complexes in the spiral arm of M~51 with 1~kpc resolution
\citep[$X_{\rm CH_3OH}$ $\sim$ 3 $\times$ 10$^{-9}$;][]{Watanabe14} as a reference of the extragalactic (i.e., kpc-scale) quiescent regions, Region~1--3 of VV~114 show a few times lower $X_{\rm CH_3OH}$, although Region 5--9 show a few times higher $X_{\rm CH_3OH}$.

\section{DISCUSSION \& SUMMARY}
\begin{figure*}
\begin{center}
\includegraphics[width=18cm]{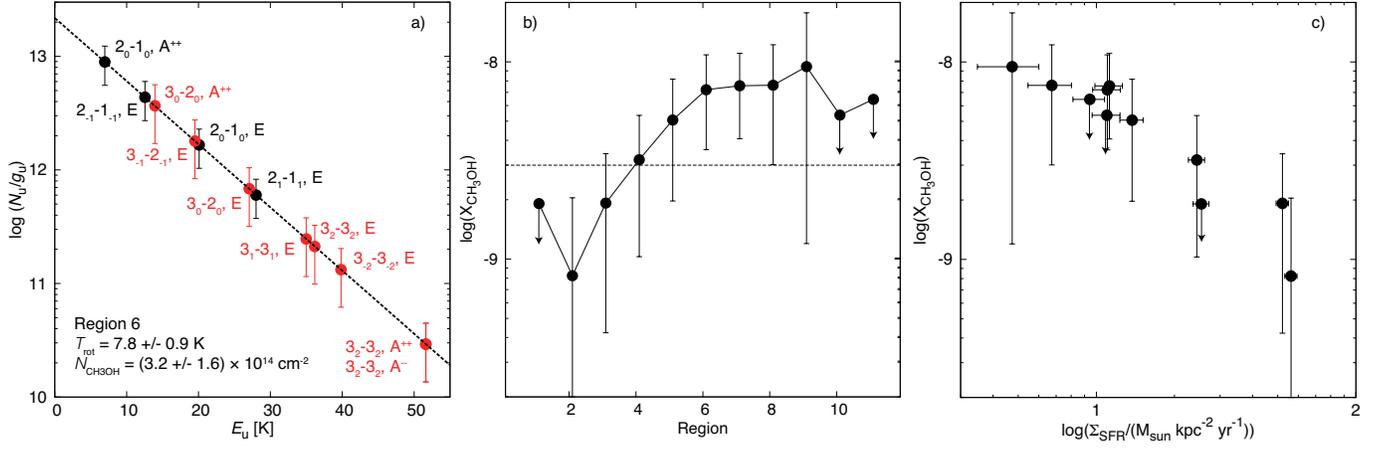}
\caption{(a) Rotation diagram of CH$_3$OH for Region 6. Black and red points show $J$ = 2--1 and 3--2 transitions, respectively. The dotted line shows the best-fit. (b) $X_{\rm CH_3OH}$ distribution along the dust filament of VV~114. The arrows show the 3$\sigma$ upper limits. The dashed line shows $X_{\rm CH_3OH}$ $\sim$ 3 $\times$ 10$^{-9}$, which is the value for the spiral arm of M51 using $\sim$ 1~kpc aperture \citep{Watanabe14}. (c) $X_{\rm CH_3OH}$ as a function of $\Sigma_{\rm SFR}$.
}
\label{fig_2}
\end{center}
\end{figure*}

To understand the characteristic CH$_3$OH distribution along the filament of VV~114, we compare $X_{\rm CH_3OH}$ with star formation rate surface density ($\Sigma_{\rm SFR}$) as shown in Figure~\ref{fig_2}c.
We employed 110~GHz continuum \citep{Saito15} assuming that free-free (bremsstrahlung) emission dominates, and applied the free-free flux to SFR conversion \citep[e.g.,][]{Yun&Carilli02}.
Such assumption is appropriate for starburst galaxies \citep[e.g., M82;][]{Condon92} and starburst-dominated LIRGs \citep[e.g., NGC~1614;][]{Saito16}.
The 110~GHz image has a similar MRS ($\sim$ 21\arcsec) and synthesized beam ($\sim$ 2\arcsec) to CH$_3$OH images.
We note that $\Sigma_{\rm SFR}$ for Region~1
and Region~2 are upper limits because of the presence of the putative AGN (i.e., contribution from nonthermal synchrotron emission).
The derived log~$\Sigma_{\rm SFR}$ shows a decreasing trend as a function of $X_{\rm CH_3OH}$ with the correlation coefficient of $-$0.94.
When we exclude a putative AGN contribution (i.e., Region~2), the correlation coefficient becomes $-$0.97.
This can be explained by efficient photodissociation of CH$_3$OH due to massive star formation and putative AGN in the nuclear region \citep[e.g.,][]{Martin09} or desorption of CO (i.e., the precursor molecule of CH$_3$OH) from dust grain surfaces into gas-phase before forming CH$_3$OH molecules due to initial high temperature conditions.
This is consistent with two orders of magnitude lower $\Sigma_{\rm SFR}$ at the spiral arm of M51 \citep{Watanabe14}.

However, the photodissociation scenario is not enough to fully explain the CH$_3$OH distribution in VV~114, because Overlap shows higher $X_{\rm CH_3OH}$ and also higher $\Sigma_{\rm SFR}$ than M51.
Thus, we need another efficient mechanism, such as hot-core or shock, in order to explain high $X_{\rm CH_3OH}$ at Overlap.
Here we estimate a possible CH$_3$OH~(2$_K$--1$_K$) flux assuming that hot-core-like environments dominate Overlap.
We used a single-dish measurement toward one of the local hot-cores NGC~2264~CMM3 (deconvolved major FWHM $\sim$ 0.076~pc, $M$ = 40~$M_{\odot}$, $I_{{\rm CH_3OH(2}_K{\rm -1}_K)}$ = 39.7~K~km~s$^{-1}$, $D$ = 738~pc), which is believed to form a massive star of 8~$M_{\sun}$ \citep[][and references therein]{Watanabe15}.
To account for all the molecular gas mass of Region~6 ($\sim$ 8.1 $\times$ 10$^8$~$M_{\odot}$) of VV~114, we need $\sim$ 2.0 $\times$ 10$^7$ CMM3.
Since the CH$_3$OH~(2$_K$--1$_K$) flux of CMM3 at 87~Mpc is $\sim$ 2.9 $\times$ 10$^{-9}$~K~km~s$^{-1}$, the total flux of CMM3-like hot-cores at Region~6 is $\sim$ 0.058~K~km~s$^{-1}$.
This is 0.5\% of the observed CH$_3$OH~(2$_K$--1$_K$) flux at Region~6 (11.9 $\pm$ 1.4~K~km~s$^{-1}$).
On average between Region~5 and Region~9, the possible contribution from hot-core-like environments is only 0.6\%.
When we use other representative hot-core environments OMC-1, Sgr~B2(N), and Sgr~B2(M) \citep{Lis93,Jones08,Bally11,Watanabe15} instead of NGC~2264~CMM3, the contribution from hot-cores at Region~6 is still small (0.7-0.9\%, 6.2\%, and 23.5\%, respectively).
We list parameters used for these comparisons in Table~\ref{table_3}.
We note that the derived parcentages are upper limits, because we assumed an extreme case that all gas masses contained in Region~6 are associated to hot-cores.
Similar estimation of a possible CH$_3$OH flux from large collections of a molecular outflow from a massive star shows a similar conclusion \citep{Meier&Turner14}.
Therefore, we suggest that kpc-scale shocks are only applicable mechanisms to explain the high $X_{\rm CH_3OH}$ at Overlap.
Possible origin of the large-scale shocks is a gas-rich galaxy-galaxy collision, because Overlap is located between the progenitor's galaxies without apparent progenitor's spiral arms, bars, or nuclei.
This shocked CH$_3$OH scenario at Overlap is consistent with the explanation of large velocity dispersion components of ionized gas detected at the same region \citep{Rich11}.
The shocked ionized gas shows the same systemic velocity as the CH$_3$OH.
This shows the evidence that ionized gas shocks co-locate with molecular gas shocks at Overlap, indicating that merger-induced shocks can affect both molecular and ionized gas ISM.

In summary, merger-induced shocks are the most likely scenario to explain the kpc-scale CH$_3$OH filament (and large dispersion components of ionized gas) across the progenitors of VV~114.
The $X_{\rm CH_3OH}$ distribution peaks at Overlap, although the eastern nucleus, which harbors dense clumps associated with a compact starburst and a putative AGN, shows almost an order of magnitude lower abundances.
$X_{\rm CH_3OH}$ clearly anticorrelates with $\Sigma_{\rm SFR}$, indicating that strong photodissociation (i.e., efficient destruction) or desorption of CO (i.e., inefficient production) on dust grains due to star-forming activities or AGN plays an important role to suppress CH$_3$OH emission at the nuclear regions.
This is the first result of merger-induced shocks in molecular gas ISM through CH$_3$OH lines.
As a future development, higher-$J$ CH$_3$OH observations are required to address more realistic, finite optical depth and non-LTE excitation conditions \citep[e.g.,][]{Goldsmith&Langer99,Mangum15}.
Avoiding the multiple $K$-ladder blending, isolated transitions (e.g., $J_K$ = 0$_0$--1$_{-1}$ at 108.89396~GHz) are also important.
Follow-up observations of shock tracers in other wavelengths \citep[e.g., near-IR warm H$_2$;][]{Sugai99,Herrera12} can be used to confirm the shock scenario at Overlap.
We will discuss other molecular lines detected and not detected in the Band~3 and Band~4 spectral scan in forthcoming paper.

\acknowledgements
The authors thank an anonymous referee for constructive comments that improved the contents of this paper.
The authors thank S. Aalto, R. Aladro, P. Sanhueza, Y. Shimajiri, and K. Sliwa for useful discussion.
T.S. and other authors thank ALMA staff for their kind support, and H. Nagai for the instruction of ALMA data reduction.
T.S. and M. Lee are financially supported by a Research Fellowship from the Japan Society for the Promotion of Science for Young Scientists.
T.S. was supported by the ALMA Japan Research Grant of NAOJ Chile Observatory, NAOJ-ALMA-0089, NAOJ-ALMA-0105, and NAOJ-ALMA-0114.
R. Kawabe was supported by JSPS KAKENHI Grant Number 15H02073.
D. Iono was supported by JSPS KAKENHI Grant Number 15H02074.
This paper makes use of the following ALMA data: ADS/JAO.ALMA\#2011.0.00467.S and ADS/JAO.ALMA\#2013.1.01057.S.
ALMA is a partner- ship of ESO (representing its member states), NSF (USA) and NINS (Japan), together with NRC (Canada), NSC and ASIAA (Taiwan), and KASI (Republic of Korea), in cooperation with the Republic of Chile. The joint ALMA Observatory is operated by ESO, AUI/NRAO, and NAOJ.

%\bibliographystyle{yahapj}

%\clearpage

\begin{deluxetable*}{cccccccccc}
\tabletypesize{\scriptsize}
\tablecaption{Information of the detected CH$_3$OH lines and their imaging properties\label{table_1}}
\tablewidth{0pt}
\tablehead{
Transition &Blended $J_K$ &$\nu_{\rm{rest}}$ &$\nu_{\rm{obs}}$ &$E_{\rm{u}}$ &S$\mu^2$ &Beam Size &Beam P.A. &$\delta v$ &Noise rms\tablenotemark{a}  \\
 & &(GHz) &(GHz) &(K) & &(\arcsec) &(\degr) &(km s$^{-1}$) &(mJy beam$^{-1}$)
}
\startdata
2$_K$--1$_K$ & 2$_{-1}$--1$_{-1}$, $E$ &\phantom{1}96.73936 &\phantom{1}94.88251 &12.5 &1.21 &1.47 $\times$ 1.12 &$-$88.8 &50 &0.37 \\
& 2$_0$--1$_0$, $A^{++}$ &\phantom{1}96.74138 &\phantom{1}94.88449 &\phantom{1}7.0 &1.62 & & & & \\
& 2$_0$--1$_0$, $E$ &\phantom{1}96.74455 &\phantom{1}94.88760 &20.1 &1.62 & & & & \\
& 2$_1$--1$_1$, $E$ &\phantom{1}96.75551 &\phantom{1}94.89835 &28.0 &1.24 & & & & \\
3$_K$--2$_K$ & 3$_0$--2$_0$, $E$ &145.09371 &142.30873 &27.1 &2.42 &1.60 $\times$ 1.04 &+67.8 &20 &0.69 \\
& 3$_{-1}$--2$_{-1}$, $E$ &145.09737 &142.31232 &19.5 &2.16 & & & & \\
& 3$_{0}$--2$_{0}$, $A^{++}$ &145.10315 &142.31799 &13.9 &2.43 & & & & \\
& 3$_{2}$--2$_{2}$, $A^{--}$ &145.12441 &142.33884 &51.6 &1.36 & & & & \\
& 3$_{2}$--2$_{2}$, $E$ &145.12619 &142.34058 &36.2 &1.33 & & & & \\
& 3$_{-2}$--2$_{-2}$, $E$ &145.12639 &142.34078 &39.8 &1.35 & & & & \\
& 3$_{1}$--2$_{1}$, $E$ &145.13186 &142.34614 &35.0 &2.21 & & & & \\
& 3$_{2}$--2$_{2}$, $A^{++}$ &145.13346 &142.34772 &51.6 &1.36 & & & &
\enddata
%% Any table notes must follow the \end{tabular} command.
\tablenotetext{a}{Noise rms in the data which have velocity resolution of $\delta v$.}
\end{deluxetable*}

\begin{deluxetable*}{lccccccccc}
\tabletypesize{\scriptsize}
\tablecaption{Information of the rotation diagram and gas and star formation properties\label{table_2}}
\tablewidth{0pt}
\tablehead{
ID & R.A. & Decl. & $S_{2_K-1_K}dv$ & $S_{3_K-2_K}dv$ & $T_{\rm rot}$\tablenotemark{a} & $N_{\rm CH_3OH}$ & $\Sigma_{\rm SFR}$\tablenotemark{b} & $N_{\rm H_2}$\tablenotemark{c} & $X_{\rm CH_3OH}$ \\
 & (\degr) & (\degr) & (Jy km s$^{-1}$) & (Jy km s$^{-1}$) & (K) & (10$^{14}$ cm$^{-2}$) & ($M_{\odot}$ kpc$^{-2}$ yr$^{-1}$) & (10$^{22}$ cm$^{-2}$) & (10$^{-9}$)
}
\startdata
1 & 16.9485 & -17.5067 & $<$ 0.26 & $<$ 0.32 & \nodata &  $<$ 0.9 & 11.9 $\pm$ 0.5 & 4.7 $\pm$ 0.4 & $<$ 1.9 \\
2 & 16.9482 & -17.5070 & $<$ 0.27 & 0.29  $\pm$ 0.08  & \nodata & 0.8 $\pm$ 1.2 & 28.2 $\pm$ 0.9 & 9.8 $\pm$ 1.0 & 0.8 $\pm$ 1.2 \\
3 & 16.9478 & -17.5072 & $<$ 0.26 & 0.68  $\pm$ 0.10  & \nodata & 1.9 $\pm$ 1.5 & 27.0 $\pm$ 0.9 & 9.8 $\pm$ 1.0 & 1.9 $\pm$ 1.5 \\
4 & 16.9474 & -17.5072 & $<$ 0.27 & 0.85  $\pm$ 0.11  & \nodata & 2.4 $\pm$ 1.6 & 14.1 $\pm$ 0.5 & 7.4 $\pm$ 0.6 & 3.2 $\pm$ 2.2 \\
5 & 16.9470 & -17.5072 & 0.44  $\pm$ 0.09  & 0.81  $\pm$ 0.11  & 6.1 $\pm$ 0.7 & 2.9 $\pm$ 1.7 & \phantom{0}9.5 $\pm$ 0.4 & 5.6 $\pm$ 0.5 & 5.1 $\pm$ 3.1 \\
6 & 16.9465 & -17.5073 & 0.51  $\pm$ 0.09  & 1.20  $\pm$ 0.12  & 7.8 $\pm$ 0.9 & 3.2 $\pm$ 1.6 & \phantom{0}7.5 $\pm$ 0.4 & 4.5 $\pm$ 0.5 & 7.2 $\pm$ 3.6 \\
7 & 16.9461 & -17.5074 & 0.57  $\pm$ 0.09  & 1.20  $\pm$ 0.11  & 7.0 $\pm$ 0.7 & 3.6 $\pm$ 1.6 & \phantom{0}6.2 $\pm$ 0.3 & 4.7 $\pm$ 0.4 & 7.6 $\pm$ 3.5 \\
8 & 16.9457 & -17.5075 & 0.33  $\pm$ 0.07  & 0.86  $\pm$ 0.11  & 8.9 $\pm$ 1.5 & 2.1 $\pm$ 1.2 & \phantom{0}4.9 $\pm$ 0.3 & 2.8 $\pm$ 0.3 & 7.6 $\pm$ 4.6 \\
9 & 16.9453 & -17.5076 & $<$ 0.26 & 0.56  $\pm$ 0.09  & \nodata & 1.6 $\pm$ 1.3 & \phantom{0}5.5 $\pm$ 0.3 & 1.6 $\pm$ 0.3 & 9.5 $\pm$ 8.3 \\
10 & 16.9449 & -17.5076 & $<$ 0.26 & $<$ 0.32 & \nodata & $<$ 0.9 & \phantom{0}6.5 $\pm$ 0.3 & 1.7 $\pm$ 0.3 & $<$ 5.4 \\
11 & 16.9444 & -17.5077 & $<$ 0.26 &$<$  0.32 & \nodata & $<$ 0.9 & \phantom{0}5.7 $\pm$ 0.3 & 1.4 $\pm$ 0.3 & $<$ 6.5
\enddata
%% Any table notes must follow the \end{tabular} command.
\tablenotetext{a}{For regions without CH$_3$OH~(2$_K$--1$_K$) detection, we adopt 7.4~K to derive $N_{\rm CH_3OH}$.}
\tablenotetext{b}{Star formation rate surface density derived from the 110~GHz continuum emission \citep{Saito15}.}
\tablenotetext{c}{Molecular hydrogen column density derived from the 880~$\mu$m dust continuum emission \citep{Saito15}.}
\end{deluxetable*}

\begin{deluxetable*}{lccccccccc}
\tabletypesize{\scriptsize}
\tablecaption{Contribution of some representative hot-cores to Region~6\label{table_3}}
\tablewidth{0pt}
\tablehead{
 & Unit & NGC~2264~CMM3 & OMC-1 & Sgr~B2(N) & Sgr~B2(M)
}
\startdata
Distance & pc & 738 & 437 & 8.34 $\times$ 10$^3$ & 8.34 $\times$ 10$^3$ &  \\
Mass & $M_{\odot}$ & 40 & 60-80 & 9100 & 2900 \\
$I_{\rm CH_3OH(2_K-1_K)}$ & K km s$^{-1}$ & 39.7 & 309 & 907 & 1114 \\
\# of hot-core at Region~6 &  & 2.0 $\times$ 10$^7$ & (1.0-1.4) $\times$ 10$^7$ & 8.9 $\times$ 10$^4$ & 2.8 $\times$ 10$^5$ \\
$I_{\rm CH_3OH(2_K-1_K)}$ at 87~Mpc & K km s$^{-1}$ & 2.9 $\times$ 10$^{-9}$ & 7.8 $\times$ 10$^{-9}$ & 8.3 $\times$ 10$^{-6}$ & 1.0 $\times$ 10$^{-5}$ \\
$I_{\rm CH_3OH(2_K-1_K)}$ from hot-cores at Region~6 & \% & 0.5 & 0.7-0.9 & 6.2 & 23.5
\enddata
%% Any table notes must follow the \end{tabular} command.
%\tablenotetext{a}{For regions without CH$_3$OH~(2$_K$--1$_K$) detection, we adopt 7.4~K to derive $N_{\rm CH_3OH}$.}
\end{deluxetable*}

\end{document}